# Original orthorhombic tetrahedral-trigonal hybrid allotropes $C_n$ (n= 8, 10, 12, 14) with *ethene*–like and *propadiene*–like units: Crystal chemistry and first principles.

Samir F. Matar
Lebanese German University (LGU), CMMS, Sahel Alma, Lebanon
https://orcid.org/0000-0001-5419-358X

**Abstract.**

*Original carbon allotropes, orthorhombic $C_8$, $C_{10}$, $C_{12}$ and $C_{14}$ presenting mixed $sp^2/sp^3$ carbon hybridizations exhibiting C=C ethene-like and C=C=C propadiene-like embedded units are proposed from crystal chemistry and calculations within the quantum density functional theory DFT. The carbon allotropes with topologies related with* ***jeb***, ***mog****, as well as new topologies, show alternating tetrahedral and trigonal carbon stacking along the a-orthorhombic direction (vertical) close to but different from "isoglitter". The carbon allotropes ($C_8$, $C_{10}$, $C_{12}$, $C_{14}$) were shown to be cohesive and stable both mechanically (elastic properties) and dynamically (phonons), with calculated Vickers hardness ($H_V$) magnitudes ranging between 25 GPa and 52 GPa, the latter magnitude assigned to $C_{12}$ possessing the largest number of tetrahedral versus trigonal stacking. High phonon frequencies $\omega$ ~50 THz were attributed to stretching mode of C=C (in $C_8$ and $C_{12}$) and C=C=C (in $C_{10}$ and $C_{14}$) components with good agreement with experiment for Raman molecular C=C intense stretching mode. Magnitudes of $\omega$ ~40 THz were proposed as signatures of C-C simple bonds in the tetrahedra. The electronic band structure and electronic density of states DOS shown exemplarily for $C_8$ point to metallic-like behavior assigned mainly to the itinerant role of trigonal carbon $\pi$-electrons.*

## Introduction

The challenging field of carbon research has a particular position among scientists, especially with diamond recognized as the hardest material. Beside this aristo-type, there is an ongoing search for original carbon allotropes that achieve similar mechanical and thermal properties with help provided by modern materials research programs based on evolutionary crystallography such as CALYPSO [1]. A library of such original mainly artificial allotropes is devoted to store the structures (SACADA) [2], classified by different topologies that are identified using TopCryst program [3]. For instance, diamond is labeled as **dia**, and its rare hexagonal form *Lonsdaleite*, is labeled along with similar or derived allotropes by **lon**.

The tetrahedral C($sp^3$) hybridization makes diamond a perfectly covalent system with large electronic band gap of ~5 eV. However, a change to semi-conductive and metallic behaviors useful for applications in electronics, can be induced by self-doping with C($sp^2$). Back in 1994, Bucknum et al. [4] proposed tetragonal $C_6$ allotrope called "glitter", i.e. shining and conductive. The crystal system was built ad hoc from 1,4-cyclohexadienoid units. Topology wise, glitter $C_6$ is called **tfi** and classified in SACADA data base under No. 95. Another arrangement of cyclohexadiene-type units was identified by us recently based on diamond-like template structure with an original tetragonal $C_6$ we called "neoglitter" characterized by ultra-hardness with metallic behavior and identified with **tfa** topology [5].

In 2012, a mixed $sp^3/sp^2$ carbon system was also identified for orthorhombic $C_8$ called "isoglitter" with **jeb** topology (actually **jeb,bbe-3,4-Cmmm** as referred to by TopCryst) and SACADA No. 105 classification [6]. The structure shown in Fig. 1a exhibits along the '*a'* orthorhombic direction (vertical axis) a succession of one trigonal layer -white central spheres- with ∠C-C-C angle of 120°corresponding to C($sp^2$) hybridization, then followed by a tetrahedral layer with C($sp^3$) (brown central sphere) with perfect C($sp^3$) tetrahedral ∠C-C-C angle of 109.47°, and lastly 2 trigonal layers (white central spheres). The relationship with organic chemistry with the use of 1,4-cyclohexadienoid cycle stands well but it also reproduces an ethene-like >C=C< unit with the white spheres aligned vertically (connected here with terminal carbons) whilst ethene or ethylene molecule has the formula $C_2H_4$. Another similar structure regarding the double tetrahedra stacks is **mog** $C_6$ (SACADA No. 96) shown in Fig. 1b. However, a major difference appears with the connection of the two tetrahedral layers by squares of carbon atoms as shown with the polyhedral projection; such configuration is deemed less stable as shown later.

Figure 1c shows our novel orthorhombic $C_8$ having **jeb** topology with another connection of two tetrahedral arrays connected through trigonal C=C (white spheres). The structure close to "isoglitter" is hence called by us "metaglitter".

Preliminary investigation of the relative ground state cohesive energies led to $E_{coh.}$/atom (metaglitter $C_8$) = -2.20 eV versus $E_{coh.}$/atom (isoglitter $C_8$)= -2.15 eV, and $E_{coh.}$/atom (**mog**-$C_6$) = -1.01 eV, which is then the least cohesive/stable.

The purpose of the present work is to delve further in the context of carbon allotropes with mixed hybridizations, by proposing novel orthorhombic allotropes with diene >C=C< units as well as larger systems embedding propadiene-like >C=C=C< units (propadiene molecule: $C_3H_4$).

All studied systems, rationalized through crystal chemistry approaches were quantitatively investigated within the quantum mechanics Density Functional Theory (DFT) [7,8], specifically their cohesiveness, electronic, mechanical, and dynamic stabilities as well as the electronic properties in relation with the presence of C=C / C=C=C units.

## 1- Computational methodology

For the devised stoichiometries the identifications of the ground state structures characterized by energy minima (and derived cohesive energies) and the prediction of their mechanical and dynamical properties called for calculations with methods based on the DFT. Herein, we used the Vienna Ab initio Simulation Package (VASP) code [9,10] and the projector augmented wave (PAW) method [10,11] for the atomic potentials. Exchange correlation (XC) effects were considered using the generalized gradient functional approximation (GGA) [12]. The relaxation of the atoms onto the ground state structures was performed with the conjugate gradient algorithm according to Press *et al.* [13]. The Blöchl tetrahedron method [14] was used for geometry optimization and energies. The Brillouin-zone (BZ) integrals were approximated by a special **k**-point sampling according to Monkhorst and Pack [15]. Structural parameters were optimized until atomic forces were below 0.02 eV/Å and all stress components were < 0.003 eV/Å$^3$. The calculations were converged at an energy cutoff of 400 eV for the plane-wave basis set in terms of the **k**-point integration in the reciprocal space from $k_x(6)\times k_y(6)\times k_z(6)$ up to $k_x(12)\times k_y(12)\times k_z(12)$ for the final convergence and relaxation to zero strains. In the post-processing of the ground state electronic structures, the charge density projections were operated on the lattice sites. The mechanical stability and hardness were obtained from the calculations of the elastic constants. The treatment of the results was done thanks to ELATE [16] online tool devoted to the analysis of the elastic tensors providing the bulk B and shear G modules along different averaging methods; Voigt's method was used herein [17]. For the calculation of the Vickers hardness a semi-empirical model based on elastic properties was used [18]. The dynamic stabilities were confirmed from the phonon positive magnitudes. The corresponding phonon band structures were obtained from a high resolution of the Brillouin Zone according to Togo *et al*. [19]. The electronic band structures were obtained using the all-electron DFT-based ASW method [20] and the GGA XC functional [12]. The VESTA (Visualization for Electronic and Structural Analysis) program [21] was used to visualize the crystal structures and charge densities.

## 2- Crystal chemistry rationale

### 2.1. $C_8$ "metaglitter" allotrope

Besides the above liminary investigations reporting on the comparative cohesive energy magnitudes between three carbon allotropes favoring presently proposed "metaglitter" $C_8$, preliminary mechanical stability criteria obtained with similar conditions of calculations showed all-positive elastic constants (*vide infra*) versus presence of negative elastic constants in "isoglitter". The "metaglitter" $C_8$ allotrope crystal parameters obtained after full geometry optimization to the ground state structure are given in Table 1a (the structure was already shown in Fig. 1c). The atoms are distributed over two four-fold Wyckoff positions, namely (4*g*) brown spheres and (4*h*) with white spheres of the base-centered orthorhombic space group *Cmmm* No. 65. The largest lattice constant is $a_{orth}$. so that the vertical orientation of the structure is along this parameter. The shortest interatomic distance of 1.34 Å is between the two trigonal carbon (white) atoms, equivalent to the magnitude admitted for C=C in ethene. This value is intermediate between d(C−C) = 1.54 Å for C($sp^3$) as in diamond, and triple bond d(C≡C) = 1.20 Å as in ethyne or acetylene ($C_2H_2$) with C($sp^1$) hybridization. The other distances are slightly below and above d(C−C). The angles show the tetrahedral configuration signature with ∠C1-C1-C1= 109.48°, while noting that the angle ∠C2-C1-C2= 114.54° is smaller but close to trigonal magnitude of 120°.

### 2.1. Addressing the problematic of stacking tetrahedra.

The larger cohesive energy of metaglitter $C_8$ versus the two other allotropes above could be assigned to the larger number of tetrahedral layers connected by trigonal carbons. The interplay between the stacking and connection of the tetrahedra and the resulting cohesive energy having been addressed, it was relevant to examine the effect of creating double layer-stacking of tetrahedra within the base-centered orthorhombic carbon sublattices by going to higher stoichiometry.

Orthorhombic $C_{12}$ was subsequently obtained by additional tetrahedral carbon positioned at a second four-fold Wyckoff position (4*h*) versus $C_8$. The geometry optimization led to a stable system with energy $E_{coh.}$/atom ($C_{12}$) = -2.26 eV (Table 1 2$^{nd}$ column), i.e. more stable than $C_8$, thus confirming the hypothesis that a larger content of tetrahedral stacks as shown in Fig. 1d, is responsible for the higher cohesive energy and the system is closer to diamond cohesive energy which amounts to -2.47 eV/atom. Concomitantly the density of $C_{12}$ increases versus $C_8$ despite the larger volume, letting confirm the role played by additional tetrahedral carbon in the structure densification. The distance between trigonal carbon (white spheres along vertical direction) amounts to 1.34 Å, as in $C_8$. The angles are of two types: trigonal for ∠C2-C2-C3=122.99° (close to 120°) and tetrahedral with ∠C1-C3-C1=109.97° and ∠C1-C1-C3=109.15°. Different topology was identified for $C_{12}$ with the label: **3,4,4T154**.

2.3. Embedding C=C=C propadiene-like units: base centered orthorhombic $C_{10}$ and $C_{14}$ allotropes.

Following the effect brought by additional tetrahedra leading to enhanced stability, it was relevant to check the effect of inserting in $C_8$ an additional trigonal carbon occupying the two-fold Wyckoff (2c) 0, ½, ½ position in space group *Cmmm*, i.e., the center of the plane orthogonal to the vertical direction (orthorhombic *a*), i.e. a face center position as shown with the red spheres in Fig. 2a. After full geometry optimization $C_{10}$ was obtained with structure parameters shown in Table 2, 1$^{st}$ column; the topology is **jeb**. Expectedly a larger volume versus $C_8$ was found due to the larger interatomic distances. Also, the angles show closer magnitudes to trigonal hybridization with 115.44° and 122.28° versus one tetrahedral-like angle ∠C2-C1-C1 of 109.16°. The additional carbon shown in red forms with the two-neighboring white (trigonal) atoms a >C=C=C< propadiene linear entity with terminal carbons instead of hydrogen in the propadiene molecule $C_3H_4$. The cohesive energy is found with lower magnitude than pristine $C_8$ with $E_{coh.}$/atom ($C_{10}$)= -1.81 eV. Then, increasing the trigonal character of the structure plays opposite role versus tetrahedral character. The density decreases due to the larger volume of $C_{10}$ on one hand and to the introduction of more trigonal carbon, on the other hand.

Lastly, $C_{14}$ was also devised based on $C_{12}$, with converged ground state described in 2$^{nd}$ column of Table 2 and depicted in Figure 2b exhibiting the propadiene units as in $C_{10}$. The extension of the lattice versus $C_{12}$ leads to a larger cohesive energy of $E_{coh.}$/atom ($C_{14}$)= -2.02 eV, versus $E_{coh.}$/atom ($C_{12}$)= -1.81 eV. Crystal analyses showed that the two allotropes presented 3,4,4**T**154 topology, different from **jeb**.

## 3- Projection of the charge density.

In view of the structure description with the presence of C=C and C=C=C units based on organic chemistry-like reasoning; further illustration of chemistry ⇔ crystal structure relationship is needed. It can be provided qualitatively by the charge density projections shown with yellow volumes around atoms displayed in Figure 3. $C_8$ (Fig. 3a) and $C_{12}$ (Fig. 3b) show the trigonal C=C with red high charge concentration at the intersection with the plane. Tricarbon containing $C_{10}$ (Fig. 3c) and $C_{14}$ (Fig. 3d) show the C=C=C propadiene entities with large concentration observed at the plane crossing in Fig. 3c representing $C_{10}$. Such concentrations of charge densities are signatures of C=C and C=C=C since less yellow volumes intensities are observed for C-C simple bonds forming the tetrahedral network.

## 4- Physical properties: results and discussions

### *3.1 Mechanical properties from the elastic constants*

The analysis of the mechanical properties was carried out with the calculation of the elastic properties by operating finite distortions of the lattice. The system is then fully described by the bulk ($B$) and the shear ($G$) moduli obtained by averaging the elastic constants. The calculated sets of elastic constants $C_{ij}$ (i and j corresponding to directions) are given in Table 2.

All $C_{ij}$ values are positive letting expect mechanically stable systems. $C_8$ and $C_{12}$ have the largest $C_{ij}$ values whereas derived $C_{10}$ and $C_{14}$ present intermediate values. Such magnitudes are translated by the bulk $B_V$ and shear $G_V$ modules obtained from ELATE program [16] devoted to the analysis of the elastic tensors. The last columns of Table 2 provide the obtained $B_V$ and $G_V$ (V subscript designated adopting Voigt averaging method (ref. [17]) showing magnitudes following the trends observed for $C_{ij}$.

Subsequently the hardness according to Vickers ($H_V$) was predicted using the model of Chen et. al. [18]:

$$H_V = 0.92\ (G/B)^{1.137}\ G^{0.708}$$

with $H_V$ possessing the same unit as $G_V$, and $B_V$, i.e., GPa.

In this equation $G/B$ is called the Pugh ratio [22] that allows distinguishing ductile behavior ($G_V/B_V < 1$) from brittle behavior ($G_V/B_V > 1$). All allotropes present $G_V/B_V$ <1 with the highest values observed for $C_{12}$ close to 1 ($G_V/B_V$ =0.91). $C_{12}$ that contains the largest amount of four tetrahedral stacks is found on the verge of brittleness, whereas $C_8$ has $G_V/B_V$ =0.82 with two tetrahedral layers within the orthorhombic cell. On the other side, $C_{10}$ and $C_{14}$ characterized by large trigonal C content have $G_V/B_V$ =0.75, lower than in $C_8$ and $C_{12}$ with a rather ductile behavior. The last column of Table 2 reflects these observations with the highest magnitude of Vickers hardness $H_V(C_{12})$ = 52 GPa pointing to super-hard behavior, larger than $H_V(C_8)$ = 43 GPa on one hand and $H_V(C_{10})$ = 32 GPa, $H_V(C_{14})$ = 36 GPa showing close magnitudes, on the other hand letting assign them average hardness behaviors. Note however that $H_V(C_{12})$ = 52 GPa remains far below diamond's which amounts to ~95 GPa cf. [5] and therein cited works.

The mechanical results agree well with the crystal chemistry description above reporting on the role played by tetrahedral stacking.

### *3.2 Dynamic and thermodynamic properties from the phonons*

An important criterion of phase dynamic stability is obtained from the phonon's properties. Phonons defined as quanta of vibrations have their energy quantized thanks to the Planck constant 'h' used in its reduced form ħ (ħ = h/2π). The phonons energy is given by E = ħω where ω is the frequency. Figure 4 depicts in four panels the phonons band structures of the four allotropes along the major lines of the orthorhombic Brillouin zone (*x*-axis). The frequency along the y-axis is expressed in units of Tera-Hertz (THz, with 1 THz = 33 $cm^{-1}$). There are 3N-3 optical modes found at higher energy than three acoustic modes that start from zero energy (ω = 0) at the Γ point that defines the center of the Brillouin Zone, up to a few Terahertz. They correspond to the lattice rigid translation modes of the crystal (two transverse and one longitudinal). The remaining bands correspond to the optic modes. All phonon frequencies are found positive letting confirm the stability of the announced allotropes from the point of view of dynamic stability. The highest frequencies in all panels are close to 50 THz, equivalent to 1650 $cm^{-1}$. This magnitude is exactly within the range of C=C vibrations with a strong signal, i.e., 1625–1680 $cm^{-1}$ in the molecular state as listed in the web-available "RAMAN Band Correlation Table". The second highest vibration is found with bands around 40 THz, signature of the tetrahedral carbon [23].

The thermodynamic properties of the new phases were calculated from the phonon frequencies using the statistical thermodynamic approach [24] on a high-precision sampling mesh in the orthorhombic Brillouin zone. The temperature dependencies of the heat capacity at constant volume ($C_v$) are shown in Figure 5 for one and two tetrahedral stack allotrope $C_8$ and $C_{12}$ in comparison with experimental $C_v$ data for diamond by Victor [25]. The calculated curves of both $C_8$ and $C_{12}$ follow closely the evolution of diamond experimental points but the curves are systematically above the experimental points. Note that $C_{12}$ which contains larger number of tetrahedra versus $C_8$ presents a curve closer to diamond. The interpretation relies here too on the fact the diamond is uniquely made of tetrahedra, whereas $C_{12}$ and $C_8$ possess trigonal and tetrahedral carbons.

### *3-3 Electronic band structures and density of states DOS*

Is so far that the main feature in all four allotropes is the concomitant presence of tetragonal and trigonal carbon sites, we consider the electronic structure of $C_8$ as their illustrative representative in Figure 6. The band structure in Fig. 6a shows the bands developing along the main directions of the orthorhombic BZ. The zero energy along the y-axis is considered with respect to the Fermi level $E_F$ crossed by finite bands, and there's no separation between the valence band VB and the conduction band CB at $E_F$. $C_8$ has then a metallic character. One can observe that without the dispersing of these bands crossing $E_F$ at R and Y points, $C_8$ would be semi-conducting or small-gap insulator. Assigning these bands can be better assessed from the site projected density of states DOS in Fig. 6b where the energy axis is now horizontal, and the DOS are expressed in $eV^{-1}$. Tetrahedral carbon (C1) DOS are shown in purple lines whereas trigonal carbon (C2) DOS are represented with blue-green lines. Within the VB, from -22 eV up to $E_F$ the DOS of the two carbon sites show resemblance in the major part, whence the bonding between the two carbon sub-structures C1 and C2. s-like states are found up to -15 eV, with s(C1) showing more localized character ($\sigma$ electrons) than smeared s(C2) with delocalized $\pi$ electrons. Most importantly one observed vanishingly small magnitude of C1 DOS which behave diamond C($sp^3$)-like at $E_F$ versus C2 DOS which behave C($sp^2$)-like. Therefore, C2 is responsible of the band crossing at $E_F$ and of the metallic character.

## 5- Conclusion

The main purpose of this paper was to highlight the effects on the physical properties of the concomitant presence of C($sp^3$)-like and C($sp^2$)-like carbons in a series of hybrid carbon allotropes. Orthorhombic $C_8$, and $C_{12}$ exhibit C=C ethene-like subunits whereas $C_{10}$ and $C_{14}$ present C=C=C (propadiene)-like subunits, all connecting tetrahedral stacks of layers. The proposed allotropes from crystal chemistry were quantitatively investigated with calculations of energies and energy-based quantities as the elastic and dynamic and electrons structures. All allotropes were shown to be cohesive and stable both mechanically (elastic properties) and dynamically (phonons), with calculated Vickers hardness (HV) magnitudes ranging between 25 GPa and 52 GPa; the latter large magnitude was assigned to $C_{12}$ containing the largest number of tetrahedral stackings. The highest phonon frequencies ω ~50 THz were attributed to the Raman stretching mode of diene C=C units while lower magnitudes ω ~40 THz were proposed as signatures of C-C bonds in the tetrahedra as in diamond. In $C_8$ and $C_{12}$ allotropes the temperature evolution of the specific heat $C_V$ follows diamond's experimental discreet points while being at higher values. The electronic band structure points to metallic-like behavior assigned mainly to the itinerant role of trigonal carbon delocalized π-electrons.

Table 1. Crystal structure parameters of the orthorhombic carbon allotropes.

a) $C_8$ (**jeb**) and $C_{12}$ (**3,4,4T154).**

| *Cmmm* No. 65 | $C_8$ | $C_{12}$ |
|---|---|---|
| *a*, | 7.8026 | 11.3788 |
| b, Å | 2.6786 | 2.5179 |
| *c*, Å | 2.5054 | 2.6231 |
| $V_{cell}$, Å$^3$ | 52.36 | 75.14 |
| *Density* g/cm$^3$ | 3.053 | 3.185 |
| *Distances d-d-* Å | d(C2-C2)= 1.34 d(C2-C1)= 1.49<br>d(C1-C1)= 1.64 | d(C3-C3)=1.34 d(C1-C1)=1.534<br>d(C3-C1)=1.60 |
| Angles (deg.) | ∠C2-C1-C2= 114.54°<br>∠C2-C1-C1= 108.19°<br>∠C1-C1-C1= 109.48° | ∠C2-C2-C3=122.99°<br>∠C1-C3-C1=109.97°<br>∠C1-C1-C3=109.15° |
| Atomic position | C1 (4g) 0.8107, ½, 0<br>C2 (4h) 0.4139, 0, ½ | C1 (4g) 0.7886, ½, 0<br>C2 (4h) 0.0589, 0, ½<br>C3 (4h) 0.1307, ½, ½ |
| E(coh.)/at. eV | -2.20 | -2.26 |

b) $C_{10}$ and $C_{14}$ (3,4,4**T**154 topology**)**

| *Cmmm* No. 65 | $C_{10}$ (**jeb**) | $C_{14}$ |
|---|---|---|
| *a*, | 10.6745 | 14.1952 |
| b, Å | 2.4833 | 2.4932 |
| *c*, Å | 2.6429 | 2.5732 |
| $V_{cell}$, Å$^3$ | 70.675 | 91.07 |
| *Density* g/cm$^3$ | 2.845 | 3.066 |
| *Distances d-d-* Å | d(C2-C3)=1.35 d(C2-C1)=1.56<br>d(C1-C1)=1.57 | d(C2-C4)=1.34 d(C2-C3)=1.51<br>d(C1-C2)=1.58 |
| Angles (deg.) | ∠C2-C1-C2=115.44°<br>∠C2-C1-C1=109.16°<br>∠C1-C2-C3=122.28° | ∠C3-C2-C4=124.13<br>∠C1-C1-C3=109.63°<br>∠C2-C3-C1=109.04° |
| Atomic position | C1 (4g) 0.7955, ½, 0<br>C2 (4h) 0.3734, 0, ½<br>C3 (2c) 0, ½, ½ | C1 (4g) 0.7811, ½, 0<br>C2 (4h) 0.0946, 0, ½<br>C3 (4h) 0.1541, ½, ½<br>C4 (2d) 0, 0, ½ |
| E(coh.)/at. eV | -1.81 | -2.02 |

E(C) = -6.6 eV. $E_{diamond}$ (coh.)/at. = -2.49 eV.; E(N)=-6.8 eV

Table 2. Elastic constants, bulk $B_V$, and shear $G_V$ moduli and hardness $H_V$ in GPa units.

| | $C_{11}/C_{22}$ | $C_{12}$ | $C_{13}/C_{23}$ | $C_{33}$ | $C_{44}$ | $C_{55}$ | $C_{66}$ | $B_V$ | $G_V$ | $H_V$ |
|---|---|---|---|---|---|---|---|---|---|---|
| $C_8$ | 1136/575 | 35 | 169/4 | 1173 | 31 | 120 | 510 | 375 | 308 | 43 |
| $C_{10}$ | 1272/846 | 114 | 76/68 | 289 | 340 | 112 | 46 | 325 | 243 | 32 |
| $C_{12}$ | 1092/1144 | 149 | 77/18 | 678 | 526 | 213 | 98 | 378 | 345 | 52 |
| $C_{14}$ | 1159/943 | 162 | 95/110 | 673 | 366 | 181 | 95 | 390 | 288 | 36 |

FIGURES

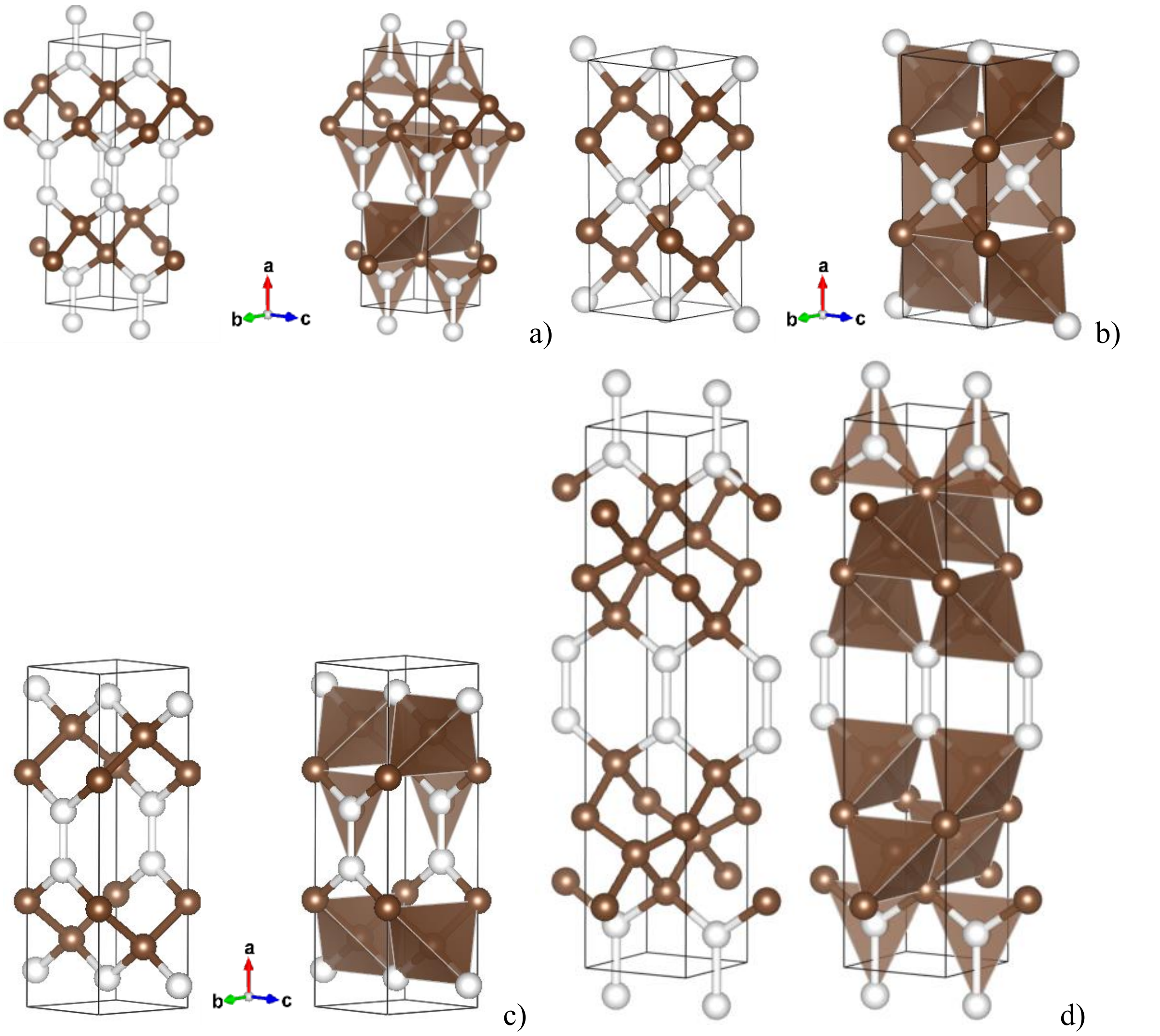

Figure 1. Sketches of the orthorhombic crystal structures with ball-and-stick (left) and polyhedral (right) representations: a) $C_8$ isoglitter (**jeb** topology) SACADA No.105, b) $C_6$ (**mog** topology) SACADA No. 96, c) $C_8$ *present work*, d) $C_{12}$ *present work.* Brown and white spheres correspond to tetrahedral and trigonal different carbon sites (cf. Table 1).

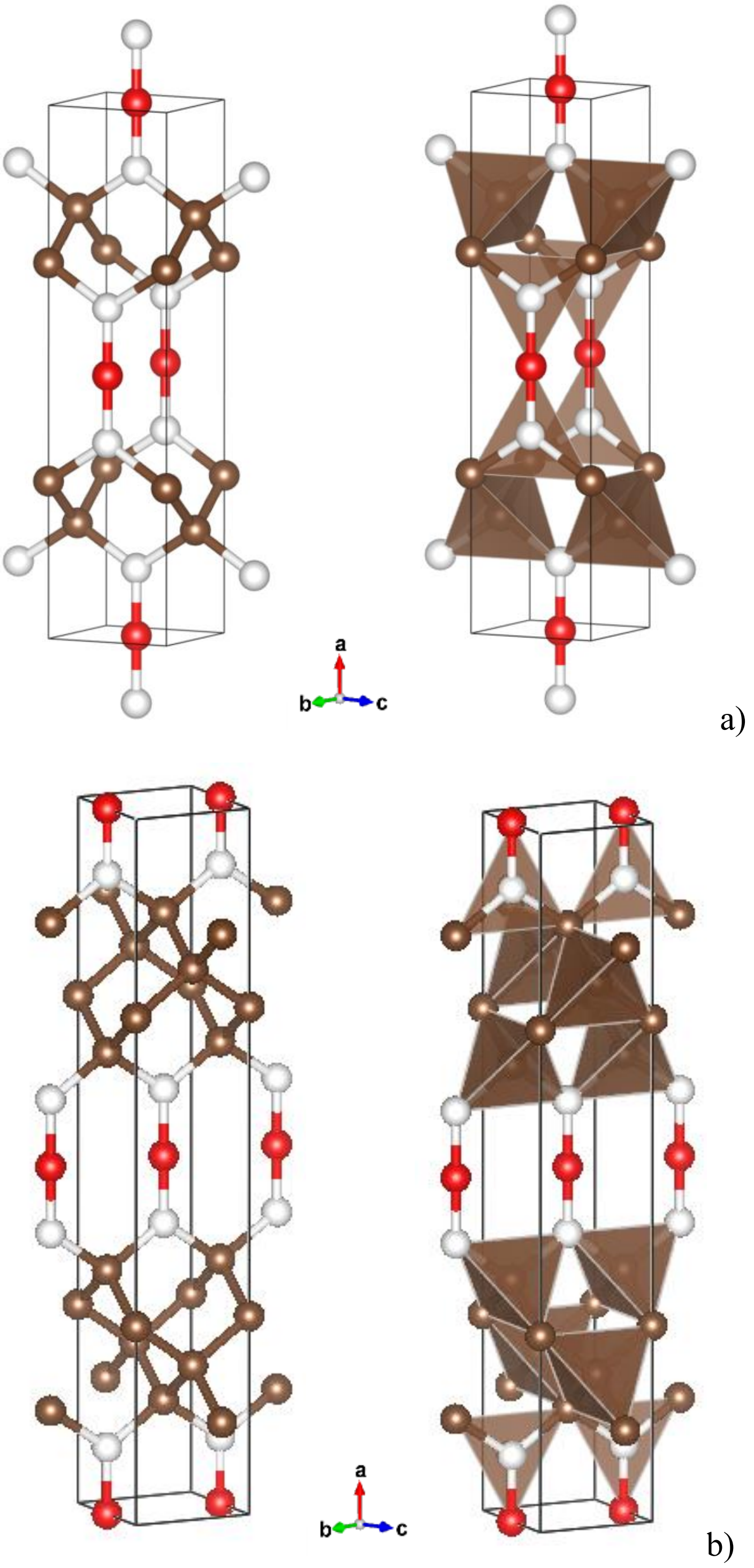

Figure 2. Sketches of the orthorhombic crystal structures with ball-and-stick (left) and polyhedral (right) representations: a) $C_{10}$, b) $C_{14}$. Brown, white and red spheres correspond to carbon on different sites (cf. Table 1).

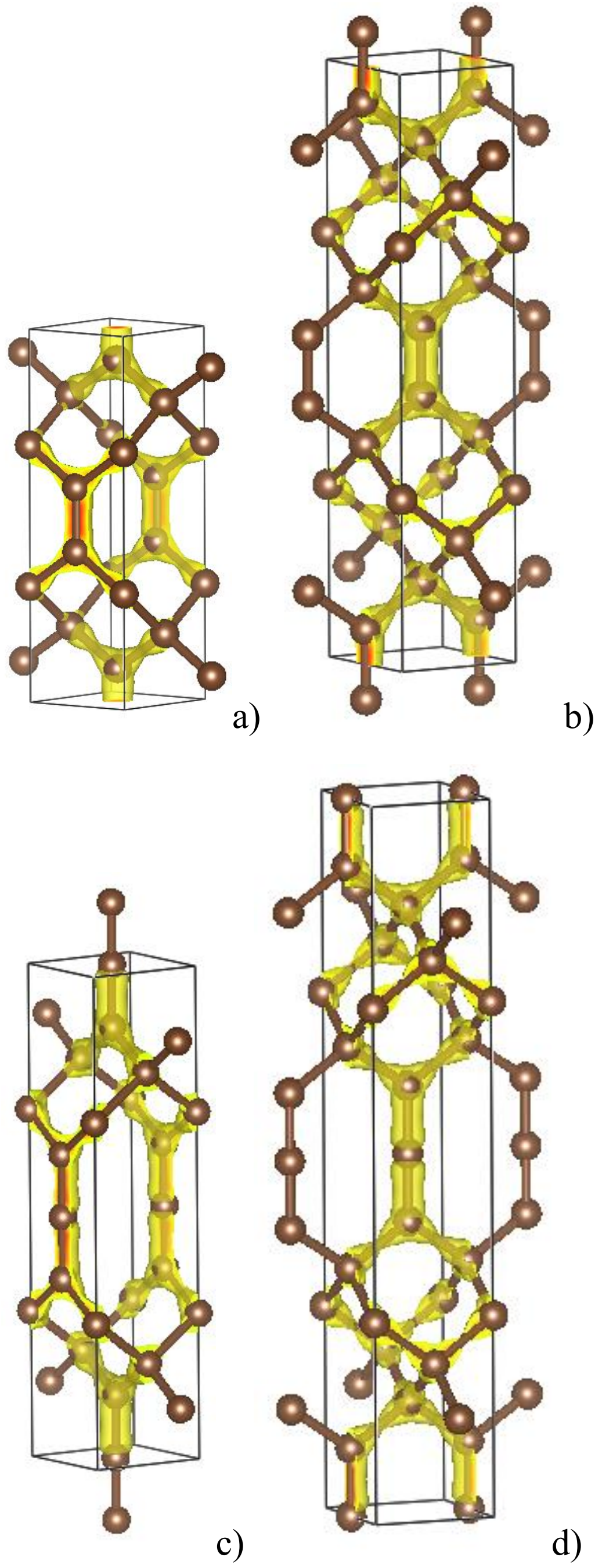

Figure 3. Charge density projections (yellow volumes) of a) $C_8$, and b) $C_{12}$, showing C=C -like diene entities, and c) $C_{10}$, and d) $C_{14}$ with propadiene C=C=C -like entities. The vertical direction is along the orthorhombic *a* lattice parameter.

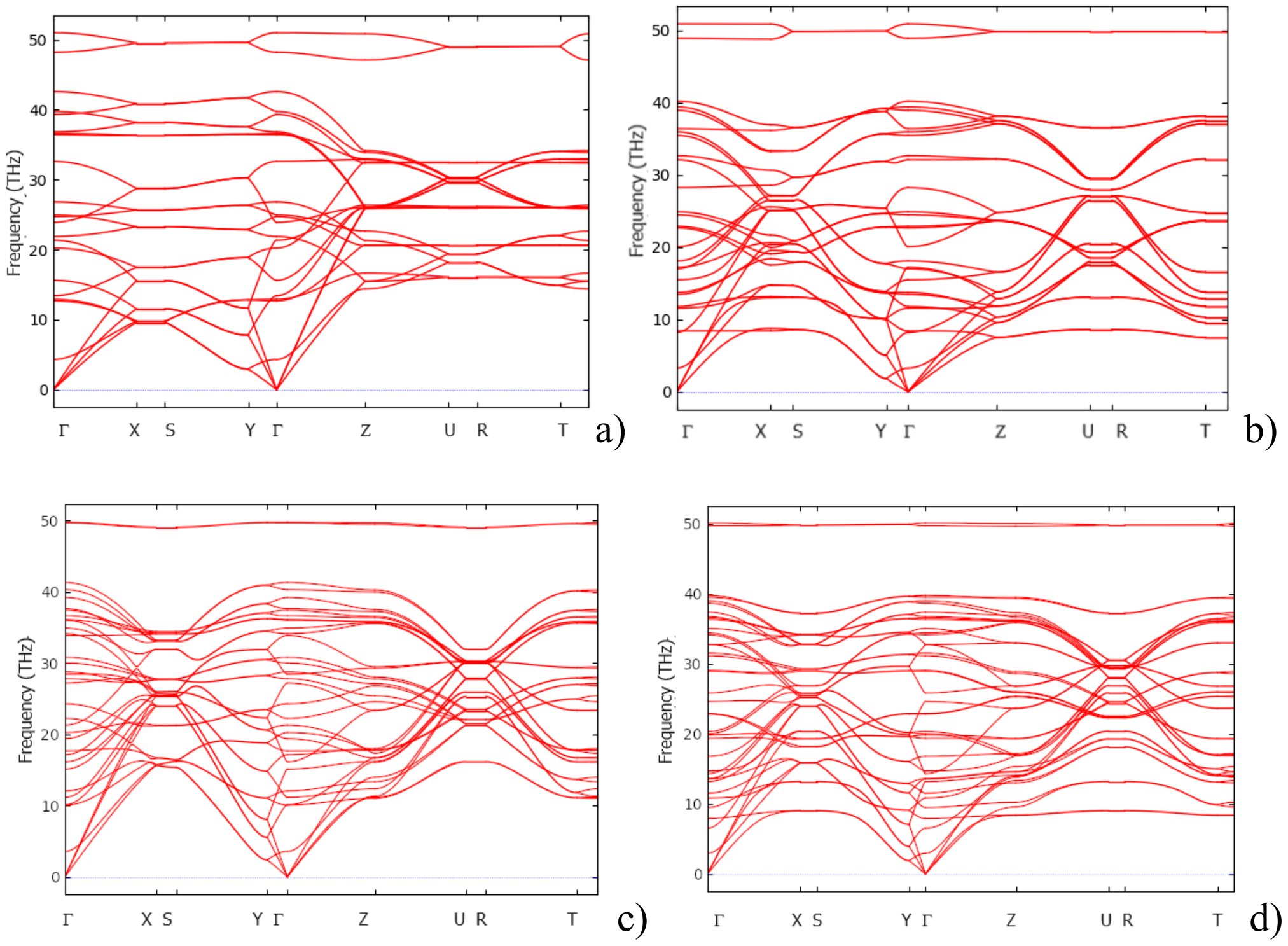

Figure 4. Phonons band structures along the major lines of the orthorhombic Brillouin zone (*x*-axis). Frequency along the y-axis is expressed in units of Terahertz (THz). a) $C_8$, b) $C_{10}$, c) $C_{12}$, and d) $C_{14}$.

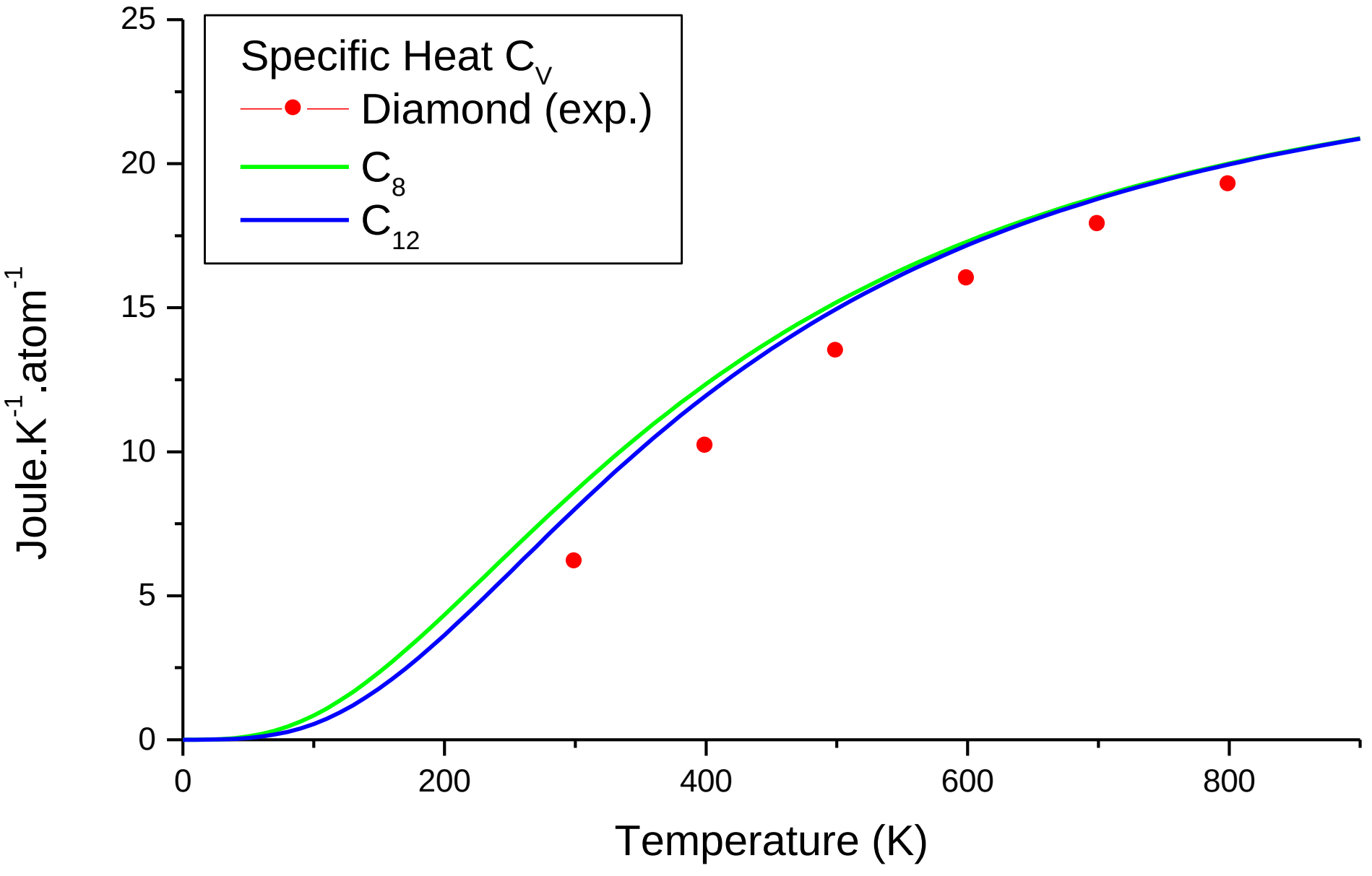

Figure 5. Thermal properties with the calculated specific heat $C_V$ of $C_8$ and $C_{12}$; the experimental values of diamond from literature are shown with red dots.

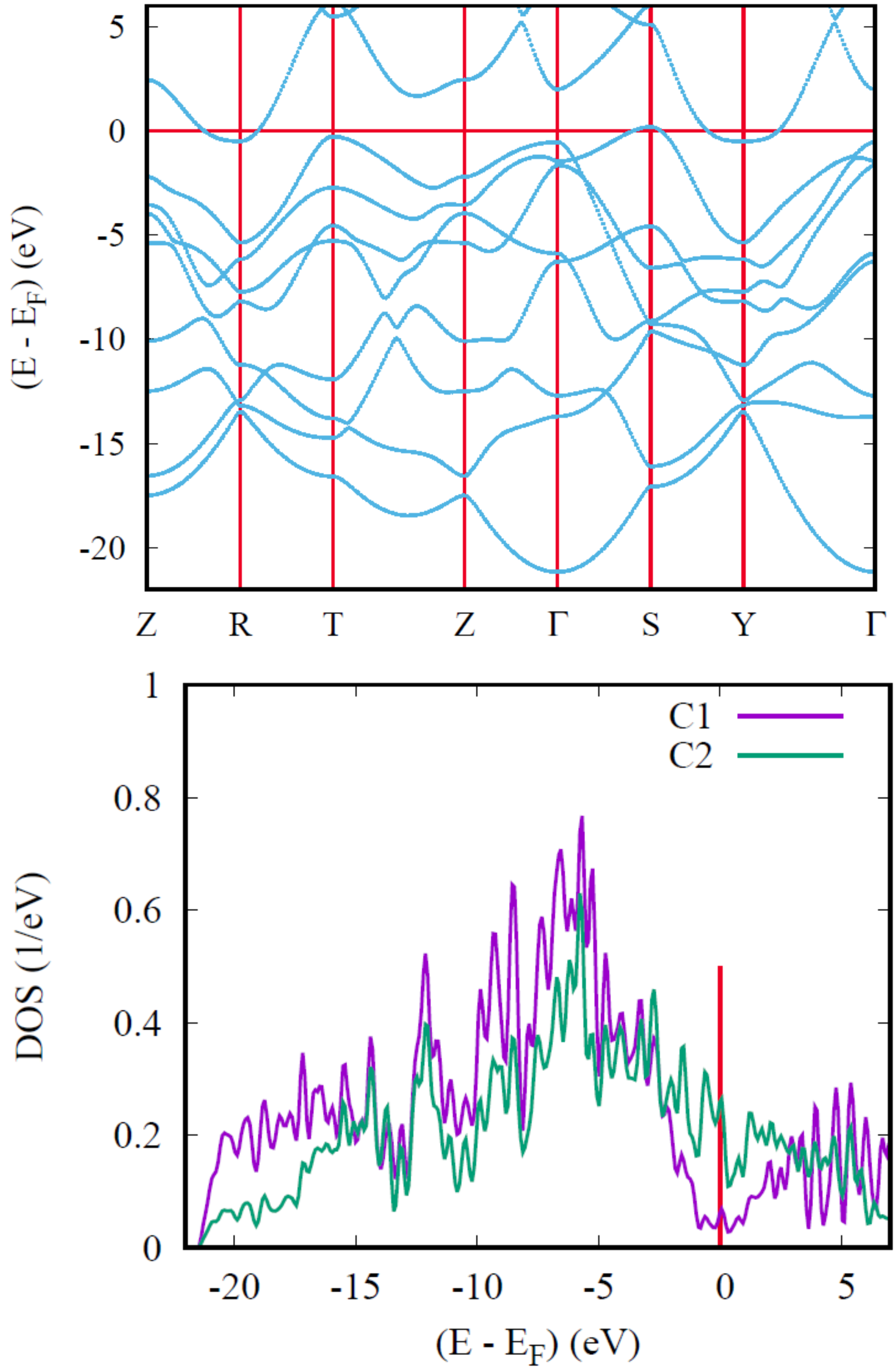

a)

b)

Figure 6. $C_8$. Electronic band structure along major lines of the orthorhombic Brillouin zone along *x*-axis (a), and site projected density of states (b). Energy is referenced with respect to the Fermi level ($E_F$) and expressed in units of electron-Volts (eV).